\documentclass[prl,aps,twocolumn,showpacs]{revtex4}
\usepackage{graphicx,color,psfrag,latexsym}
\usepackage{dcolumn}
\usepackage{amsmath,amssymb,epsf,bm}

\begin{document}

\title{Inverse Spin Hall Effect in SNS Josephson Junctions}
\author{A.~G. Mal'shukov$^{1}$, Severin Sadjina$^{2}$, and Arne Brataas$^{2}$%
}
\affiliation{$^1$Institute of Spectroscopy, Russian Academy of Sciences, 142190, Troitsk,
Moscow oblast, Russia \\
$^2$Department of Physics, Norwegian University of Science and Technology,
NO-7491 Trondheim, Norway}

\begin{abstract}
We consider DC supercurrents in SNS junctions. Spin-orbit coupling
in combination with Zeeman fields can induce an effective vector
potential in the normal conductor. As a consequence, an out-of-plane
spin-density varying along the transverse direction causes a
longitudinal phase difference between the superconducting terminals.
The resulting equilibrium phase coherent supercurrent is analogue to
the non-equilibrium inverse spin Hall effect in normal conductors.
We explicitly compute the effect for the Rashba spin orbit coupling
in a disordered two-dimensional electron gas with an inhomogeneous
perpendicular Zeeman field.
\end{abstract}

\maketitle

The spin-Hall effect (SHE) and inverse SHE (ISHE) are remarkable
demonstrations of the influence of the spin-orbit coupling on
electron transport. Via this coupling, a longitudinal electric
current can induce a perpendicular spin current and vice versa.
These effects take place in metals and semiconductors, where the
spin-orbit interaction (SOI) arises from impurity scattering
\cite{Dyakonov} or band structure effects \cite{Sinova}. Utilizing spin
injection, SHE, and ISHE, electron spins can be controlled,
as recently demonstrated experimentally \cite{Valenzuela}.

We discuss the intrinsic SHE and ISHE, where the dominant spin-orbit
coupling is from the electron band structure. The study of SHE has
been focused on normal conductors, \textit{e.g.}\ normal metals and
semiconductors. New, interesting, and rich physics occurs in
superconductors where electron transport is dissipationless and the
ground state exhibits macroscopic coherence. Some superconductivity
induced features
of the intrinsic SHE has recently been analyzed in bulk superconductors \cite%
{Contani} and SNS Josephson junctions \cite{MalshukovSNS}. The
latter work revealed an equilibrium spin accumulation at lateral
sample edges, similar to non-equilibrium spin accumulation in normal
conductors, but the spin Hall current vanished due to time-reversal
symmetry in the DC Josephson effect.

We focus on ISHE in Josephson junctions. There are two scenarios
depending on how the spin current (density) is created in the normal
metal. In a dissipative setup, additional normal/ferromagnetic
terminals in the transverse direction inject a non-equilibrium spin
current. Subsequently, the ISHE induces an electric potential
difference $V_{SH}$ between superconducting terminals, causing
Josephson oscillations at frequency $2eV_{SH}/\hbar$. Transport is
dissipative due to the spin flow between transverse
normal/ferromagnetic terminals. This phenomenon is interesting from
an experimental point of view and we will study it quantitatively
elsewhere, but we consider here a dissipationless effect.

We predict a novel inverse dissipationless SHE: An out-of-plane
equilibrium spin density spatially varying in the transverse
direction induces a longitudinal electric supercurrent.
Equivalently, it induces a phase shift between two superconducting
terminals. In general, since the equilibrium spin-density controls
ISHE, Zeeman interaction from magnetic or exchange fields
manipulates the resulting Josephson supercurrent. As an explicit
illustration, we consider the interplay of spin-orbit coupling and
Zeeman fields in a disordered two-dimensional electron gas (2DEG),
and compute the magnitude of the equilibrium Josephson ISHE.

The interplay of Zeeman field and SOI leading to
an effective phase difference between superconducting terminals
has recently also been studied in two quite different systems,
but neither exhibits the ISHE we discuss: A supercurrent in response to a \textit{%
spatially homogenous} magnetic field has been predicted for
Josephson tunneling through a 1D wire \cite{Krive} and appears in
numerical simulations of the superconducting transport through a
ballistic point contact \cite{Reynoso} in a \textit{spatially
homogenous} parallel magnetic field. In additon to our main finding
of an inverse SHE, we provide an improved understanding of these
phenomena by showing how the interplay of Zeeman field and SOI can
result in the appearance of an effective electromagnetic vector
potential. Such a vector potential, in direct analogy with the
Meissner effect, gives rise to a supercurrent.

Let us outline our model. The spin-orbit interaction arises from the band
structure, $H_{so}=\bm{\sigma}\cdot \mathbf{h}_{\bm{k}}$, where $\bm{\sigma}%
=(\sigma _{x},\sigma _{y},\sigma _{z}$) is a vector of Pauli
matrices. We ussume that the spin-orbit field $\mathbf{h}_{\bm{k}}$
is given by Rashba SOI where $h_{x}=\alpha k_{y}$ and $%
h_{y}=-\alpha k_{x}$. Two examples of spin density manipulations in
2DEG will be considered: i) a perpendicular to 2DEG Zeeman field
spatially varying in the transverse direction $y$, as shown in Fig.
1 and ii) homogeneous Zeeman field directed along the $y$-axis. We
will show that setup i) exhibits an equilibrium inverse SHE. Setup
ii) also changes the current-phase relation in SNS contacts.
\begin{figure}[tp]
\includegraphics[width=6cm]{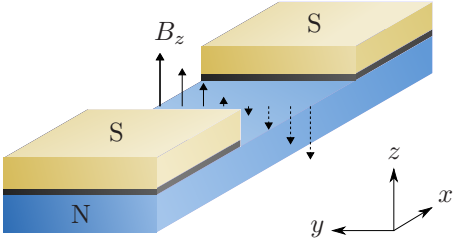}
\caption{A SNS Josephson junction. Interplay between Rashba
spin-orbit interaction and Zeeman splitting in a normal 2D film (N)
induces a phase difference between order parameters of two
superconducting terminals (S). An inhomogenous Zeeman interaction
can be created by \textit{e.g.}\ a ferromagnetic layer on top of the
film or magnetic impurities. Another possible configuration (not
shown) is a uniform field parallel to the $y$-axis} \label{fig1}
\end{figure}
All relevant length scales are assumed larger than the mean free path $%
l=v_{F}\tau $, and we are in the metallic regime $k_{F}l\gg 1$,
where $k_{F}$ and $v_{F}$ are the Fermi wave-vector and velocity,
respectively. These conditions allow a diffusion approximation in
the description of electron transport. In this regime, the transport
properties are described by a generalized Usadel equation, which we
will now derive. The resulting Usadel equation is similar to the one
in Ref. \cite{MalshukovSNS}, but important non-trivial new terms
essential for the effects we discuss are added due to the Zeeman
interaction $H_{Z}(\mathbf{r})=\sigma _{z}H_{z}(\mathbf{r})+\sigma _{y}H_{y}(%
\mathbf{r}), $ where $H_{z}$ ($H_{y}$) are the perpendicular
(in-plane) components of the Zeeman field.  We start from the
anomalous retarded thermal
equilibrium Green function $F_{\alpha \beta }(\mathbf{r},\mathbf{k},\omega )$%
, which is the Fourier transform of
\begin{equation}
F_{\alpha \beta }=-i\langle \lbrack \psi _{\alpha }(\mathbf{r}+\frac{%
\bm{\rho}}{2},t),\psi _{\beta }(\mathbf{r}-\frac{\bm{\rho}}{2},t^{\prime
}]_{+}\rangle \theta (t-t^{\prime })  \label{Frrho}
\end{equation}%
with respect to the relative coordinate $\bm{\rho}$ and relative time $%
t-t^{\prime }$.   It is convenient to use a singlet-triplet basis
representing the Green function,
\begin{equation}
F_{\alpha \overline{\beta }}=\frac{1}{\sqrt{2}}\left( \delta _{\alpha \beta
}F_{0}+\sigma _{\alpha \beta }^{z}F_{\mathrm{s}}\right) +\frac{\sigma
_{\alpha \beta }^{+}}{2}F_{1}+\frac{\sigma _{\alpha \beta }^{-}}{2}F_{-1},
\label{triplet}
\end{equation}%
where $\overline{\beta }$ denotes a spin projection opposite to $\beta $, $%
\sigma _{\alpha \beta }^{\pm }=\sigma _{\alpha \beta }^{x}\pm i\sigma
_{\alpha \beta }^{y}$. $F_{s}$ denotes the singlet component. $F_{0}$ and $%
F_{\pm 1}$ are triplet components corresponding to $0$ and $\pm 1$
projections of the Cooper's pair total spin on the $z$-axis. Using
a standard method starting from Gor'kov equations \cite{Efetov},
we derive the semiclassical equation
\begin{equation}
\sum_{m}(\delta _{nm}-i\tau K_{nm})F_{m}=\frac{i}{2\pi N_{F}}[G_{11}^{0}\Psi
+\Psi G_{22}^{0}]_{n}\,,  \label{Schrodinger}
\end{equation}%
where subscripts $n$ and $m$ attain the values $0,\pm 1$, or $\mathrm{s}$,
and
\begin{equation}
K=2\omega -\mathbf{v}\hat{\mathbf{q}}-2\mathbf{J}\mathbf{h}_{\mathbf{k}}-S-B.
\label{K}
\end{equation}%
Here $\hat{\mathbf{q}}=-i\nabla $, $\mathbf{J}$ is the 3$\times $3 matrix
spin $1$ operator in the triplet subspace, and operators $S$ and $B$ provide
mixing of triplet and singlet components:
\begin{eqnarray}
S_{\pm 1,\mathrm{s}} &=&-S_{\mathrm{s},\mp 1}= \mp\frac{\mathbf{\hat{q}}}{\sqrt{2%
}}\frac{\partial h_{\mathbf{k}}^{\mp }}{\partial \mathbf{k}}\,;\,B_{0,%
\mathrm{s}}=B_{\mathrm{s},0}=2H_{z}  \notag  \label{MB} \\
B_{\pm 1,\mathrm{s}} &=&-B_{\mathrm{s},\pm 1}=i\sqrt{2}H_{y}\,,
\end{eqnarray}%
where $h^{\pm }=h^{x}\pm ih^{y}$. In the right-hand side of Eq. (\ref%
{Schrodinger}) $\Psi =\sum_{\mathbf{k}}F$ and the unperturbed
retarded Green functions are
\begin{equation}
G_{11/22}^{0}=\left( \omega \mp E_{k}-\bm{\sigma}\cdot \mathbf{h}_{\mathbf{k}%
}-\sigma _{z}H_{z}\mp \sigma _{y}H_{y}+i\Gamma \right) ^{-1}\,.  \label{G11}
\end{equation}

The diffusion equation can be derived from Eq. (\ref{Schrodinger}) by
expanding the operator $(1-i\tau K)^{-1}$ for small $\tau K$ and averaging
over $\mathbf{k}$. The resulting Usadel equation is
\begin{equation}
2i\omega \Psi =\tau \langle \left( -i\bm{v}\cdot \frac{\partial }{\partial %
\bm{r}}+2\mathbf{J}\cdot \mathbf{h}_{\bm{k}}\right) ^{2}\rangle _{F}\Psi
-M\Psi \,,  \label{Usadel}
\end{equation}%
where the angular brackets denote averaging over the Fermi surface. The
matrix $M$ originates from the SOI and the Zeeman interaction expressed via
the operators $S$ and $B$. Its off-diagonal terms describe singlet-triplet
transitions.  The relevant matrix elements for our further analysis are :
\begin{eqnarray}  \label{Mz}
M_{\mathrm{s}\mathrm{s}} &=&2\tau ^{3}\sum_{\nu =\pm }\nu \langle b_{\mathbf{%
\hat{q}}}^{-\nu }H_{z}a_{\mathbf{\hat{q}}}^{\nu }+a_{\mathbf{\hat{q}}}^{-\nu
}H_{z}b_{\mathbf{\hat{q}}}^{\nu }\rangle _{F},  \notag  \\
M_{\mathrm{s}\pm 1} &=&\frac{4i\tau ^{2}}{\sqrt{2}}\langle 2H_{z}b_{\mathbf{%
\hat{q}}}^{\pm }+b_{\mathbf{\hat{q}}}^{\pm }H_{z}\mp \frac{1}{2}h_{\mathbf{k}%
}^{2}a_{\mathbf{\hat{q}}}^{\pm }\rangle _{F}-\sqrt{2}H_{y},  \notag \\
M_{\pm 1\mathrm{s}} &=&\frac{4i\tau ^{2}}{\sqrt{2}}\langle H_{z}b_{\mathbf{%
\hat{q}}}^{\mp }+2b_{\mathbf{\hat{q}}}^{\mp }H_{z}\mp \frac{1}{2}h_{\mathbf{k%
}}^{2}a_{\mathbf{\hat{q}}}^{\mp }\rangle _{F}+\sqrt{2}H_{y},  \notag \\
M_{\mathrm{s}0} &=&M_{0\mathrm{s}}=-2iH_{z}\,,
\end{eqnarray}%
where $a_{\mathbf{\hat{q}}}^{\pm }=\hat{q}^{i}\partial
h_{\mathbf{k}}^{\pm }/\partial \hat{k}^{i}$ and
$b_{\mathbf{\hat{q}}}^{\pm }=h_{\mathbf{k}}^{\pm }(\mathbf{v}\cdot
\mathbf{\hat{q}})$ so that \textit{e.g.} the singlet-singlet
diagonal element is proportional to $\nabla _{x}$.

In order to understand some of the underlying physics described by
Eq. (\ref{Usadel}), we will demonstrate that SOI in combination with
the Zeeman field gives rise to an effective Meissner effect. Let us
first discuss this in the most transparent "local" approximation
when the SOI is strong enough/the system long enough, so that the
spin diffusion length $L_{so}=v_{F}/\sqrt{D/\Gamma _{so}}\ll
L,\sqrt{D/T} $, where $L$ is the length of the junction, $\Gamma
_{so}=2\tau \langle h^{2}\rangle _{F}$ is the spin relaxation rate
and $D=v^2_{F}\tau /2$ is the diffusion constant. In this
approximation derivatives in triplet parts of Eq. (\ref{Usadel}) can
be disregarded, except in a narrow range $\sim L_{so}$ near the
boundaries. $H_{z}$ is assumed to vary slowly on the $L_{so}$ scale.
Expressing the triplet components of $\Psi $ via the singlet $\Psi
_{s}$ and substituting them into the singlet projection of
Eq.(\ref{Usadel}), the latter takes the form
\begin{equation}
2i\omega \Psi _{\mathrm{s}}=-D\frac{\partial }{\partial x^{2}}\Psi _{\mathrm{%
s}}+2iA\frac{\partial }{\partial x}\Psi _{\mathrm{s}}\,,
\label{Usadel2}
\end{equation}%
where $A$ is a real coefficient obtained from the equation
\begin{equation}
2iA\nabla _{x}=M_{\mathrm{ss}}+\frac{1}{\Gamma _{so}}\sum_{m=\pm 1}M_{%
\mathrm{s}m}M_{m\mathrm{s}}.  \label{A}
\end{equation}%
Here we have only included dominant terms proportional to $\alpha
^{2}\partial H_{z}/\partial y$ and $\alpha H_{y}$, where $\alpha $
is the SOI coupling constant. Higher order contributions to (\ref%
{Usadel2}) proportional to $H^{2}$ and $\alpha ^{4}$ have been disregarded.

The diffusion equation (\ref{Usadel2}) demonstrates that $cA/e$ is
an effective weak electromagnetic vector-potential. Therefore,
similar to the Meissner
effect it will induce a supercurrent. To order $A^{2}$ the solution of Eq. (%
\ref{Usadel2}) is $\Psi_{\mathrm{s}}=\Psi_{\mathrm{s}}^{0}\exp
(ixA/D)$, where $\Psi_{\mathrm{s}}^{0}$ satisfies Eq.
(\ref{Usadel2}) with $A=0$. The exponential factor gives rise to an
additional phase difference $\theta =LA/D $ between the
superconducting terminals, the Josephson current is $j_{c}\sin (\phi
+\theta )$, where $\phi $ is the initial phase difference between
the terminals and $j_{c}$ is the critical current determined by the
function $\Psi_{\mathrm{s}}^{0}$. The coefficient $A$ is simple for
Rashba SOI. For a parallel Zeeman field $A=4\alpha \tau H_{y}$. For
a perpendicular field it vanishes, which is expected since it is
similar to the behavior of the spin Hall conductance. Continuing
such an analogy, one can expect that $A\neq 0$ for the cubic
Dresselhaus \cite{Dresselhaus} SOI \cite{MalshDress}.

In order to find a finite ISHE even for the Rashba SOI, we must
extend our consideration beyond the local approximation. In this
case the diffusion equation (\ref{Usadel}) cannot be reduced to the
simple form (\ref{Usadel2}). We consider superconducting leads with
equal real order parameters $\Delta $ connected via two SN
interfaces wit a low transparency $t$. The barriers are assumed to
extend into the 2DEG under the superconducting leads, so that the
range of a free electron motion is between $x_{L}$ and $x_{R}$ at
the left and right leads, respectively. Depending on contact
fabrication, other models can be similarly studied. For example, the
electrons in the 2DEG could move freely under contacts, with the
barriers present only in $z$-direction, as shown in Fig.1. The
choice of the model is not important for the main qualitative
results obtained below.

To the lowest order in the tunneling transparency $t$, the
superconducting current can be expressed \cite{Aslamazov} as a sum
over Matsubara frequencies $\omega =\pi (2n+1)T$:
\begin{equation}
j=\frac{\pi eT}{2R_{b}^{2}N_{F}}\sum_{\omega }\frac{\Delta ^{2}}{\Delta
^{2}+\omega ^{2}}\text{Im}[\int dydy^{\prime }f_{\mathrm{ss}}(\mathbf{r}_{L},%
\mathbf{r}_{R}^{\prime })]\,,  \label{current}
\end{equation}%
where $R_{b}$ is the boundary resistance \cite{Kupriyanov}, $\mathbf{r}%
_{L/R}=(x_{L/R},y)$ and
$f_{ab}(\mathbf{r}_{L},\mathbf{r}_{R}^{\prime })$, with $a,b=0,\pm
1,\mathrm{s}$, is the Green function of Eq. (\ref{Usadel}), i.e. a
solution of Eq. (\ref{Usadel}) with a delta source in its
right-hand side. The equations for retarded and advanced functions
must be properly continued to the upper and lower complex
semiplanes of $\omega $, respectively. Treating $M$ in
(\ref{Usadel}) perturbatively one can express the correction to
$f^{(0)}_{\mathrm{ss}}(\mathbf{r}_L,\mathbf{r}_R^{\prime})$ as
\begin{eqnarray}  \label{correct}
&&\delta f_{\mathrm{ss}}(\mathbf{r}_L,\mathbf{r}_R^{\prime})=-\int d \mathbf{%
r} f^{(0)}_{\mathrm{ss}}(\mathbf{r}_L,\mathbf{r}) M_{\mathrm{s}\mathrm{s}%
}f^{(0)}_{\mathrm{ss}}(\mathbf{r},\mathbf{r}_R)+  \notag \\
&&\sum_{mm^{\prime }}\int d \mathbf{r}_1 d \mathbf{r}_2f^{(0)}_{\mathrm{ss}}(%
\mathbf{r}_L,\mathbf{r}_1)M_{\mathrm{s}m} \times \notag \\
&& f^{(0)}_{mm^{\prime }}(\mathbf{r}_1,\mathbf{r}_2)M_{m^{\prime }\mathrm{s}%
}f^{(0)}_{\mathrm{ss}}(\mathbf{r}_2,\mathbf{r}_R^{\prime})\,,
\end{eqnarray}
where the unperturbed diffusion propagators $f^{(0)}_{\mathrm{ss}}(\mathbf{r}%
,\mathbf{r}^{\prime})$ and $f^{(0)}_{mm^{\prime}}(\mathbf{r},\mathbf{r}%
^{\prime})$, with $m,m^{\prime}=0$ or $\pm 1$, are obtained from Eq. (\ref%
{Usadel}) with hard-wall boundary conditions, $\nabla_x
f^{(0)}_{\mathrm{ss}} \rightarrow 0$ at $x=x_L$ and $x=x_R$, while the
triplet components in the case of Rashba SOI satisfy the boundary condition $%
(iL_{so}\nabla_x+2J_y)f=0$ \cite{MalshAccumulation}.

To illustrate the ISHE, we consider the case of Rashba SOI
with finite $H_z$, $H_{y}=0,$ and $\Delta \gg \omega $. The
parameter of interest is the effective phase difference
\begin{equation}
\theta _{\text{eff}}=\frac{\sum_{\omega }\frac{\Delta ^{2}}{\Delta
^{2}+\omega ^{2}}\text{Im}[\int dydy^{\prime }f_{\mathrm{ss}}(\mathbf{r}_{L},%
\mathbf{r}_{R}^{\prime })]}{\sum_{\omega }\frac{\Delta ^{2}}{\Delta
^{2}+\omega ^{2}}\text{Re}[\int dydy^{\prime }f_{\mathrm{ss}}^{0}(\mathbf{r}%
_{L},\mathbf{r}_{R}^{\prime })]}  \label{phase}
\end{equation}%
From Eq. (\ref{Usadel}), $\theta _{\text{eff}}=C\phi $, where $C=2l\overline{%
\nabla _{y}H_{z}}\tau /k_{F}L$ and $\overline{\nabla _{y}H_{z}}$ denotes the
average value of the the magnetic field gradient in the contact range. $\phi
$ is shown in Fig. 2 as a function of ratio of the spin relaxation rate $%
\Gamma _{so}=2\tau \alpha ^{2}k_{F}^{2}$ versus the Thouless energy $%
E_{T}=D/L^{2}$, a convenient measure of the SOI strength. For large SOI the
"local" approximation is obtained by using in the second term of Eq. (\ref%
{correct}) the approximate form of $f_{\pm 1\pm 1}=2f_{00}=-\delta (\mathbf{r%
}_{1}-\mathbf{r}_{2})/\Gamma _{so}$. In this case both terms in (\ref%
{correct}) are proportional to $\alpha ^{2}$ and precisely cancel
each other, as in the factor $A$ in (\ref{Usadel2}). Beyond this
leading "local" approximation there are terms increasing slower
than $\alpha ^{2}$. They contribute to Fig. 2.
\begin{figure}[bp]
\includegraphics[width=6cm, height=6cm]{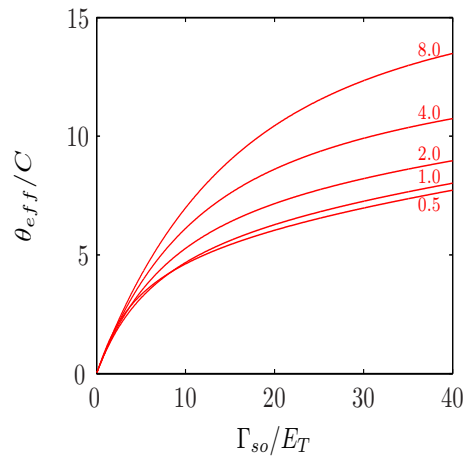}
\caption{The phase difference versus a ratio of the spin relaxation
rate and the Thouless energy, at 0.5$< k_B T/E_T < 8$. The parameter
$C$ is described in the text.} \label{fig2}
\end{figure}

Larger Zeeman fields cannot be treated perturbatively. A strong depairing
effect takes place when the characteristic length $L_{Z}=\sqrt{D/2H}$ is
small, $L_{Z}\ll $ min$(L,L_{so})$. Then, for $H=H_{z}$, both $\Psi_{\mathrm{%
s}}$ and $\Psi_{0}$ decay exponentially near contacts with superconducting
terminals and the latter become effectively disconnected. On the other hand,
as it follows from Eq. (\ref{Usadel}), $\Psi_{\pm 1}$ components are not
subject to the depairing effect and can propagate at the relatively large
distance $\sim L_{so}$. Such a long-range triplet effect has been studied in
SFS junctions, where a link between triplet and singlet Cooper pairs has
been induced by an inhomogeneous (rotating) magnetization (see \cite{Efetov}
and references therein). In our case, a coupling of $\Psi_{\pm 1}$ to $%
\Psi_{0}$ and $\Psi_{\mathrm{s}}$ can be provided by SOI through the matrix
elements $M_{\pm 1\mathrm{s}}$ and the spin precession operator $R_{\pm
1,0}=-i4\tau \langle (\mathbf{J}_{\pm 1,0}\cdot \mathbf{h}_{\bm{k}})(\bm{v}%
\cdot \frac{\partial }{\partial \bm{r}})\rangle _{F}$ originating
from the first term in the right-hand side of Eq. (\ref{Usadel}).
Indeed, assuming that $H_{z}\gg \Gamma _{so}$ and $E_{T}$, it is
easy to show that the modified Eq. (\ref{correct}) is represented
by its second term, where the integrand has the form
\begin{equation}
f_{\mathrm{s0}}^{(0)}(x_{L},x)R_{0m}f_{mm}^{(0)}(x,x^{\prime })M_{m\mathrm{s}%
}f_{\mathrm{ss}}^{(0)}(x^{\prime },x_{R}).  \label{integrand}
\end{equation}%
The unperturbed functions $f_{\mathrm{s0}}^{(0)}$ and $f_{\mathrm{ss}}^{(0)}$
are obtained from $s$ and $0$ projections of Eq. (\ref{Usadel}), where the
precession term and all $M_{ij}$, except $M_{\mathrm{s}0}$ and $M_{0\mathrm{s%
}}$ are ignored. The physics of the process described by
(\ref{integrand}) is clear: the magnetic field mixes 0-triplet and
singlet components of the pairing function within the short range
near the left boundary. Further, due to the spin precession in the
SOI field the 0-triplet transforms to $\pm 1$ triplet components.
The latter propagate to the right contact where they convert to the
singlet through $M_{\pm 1\mathrm{s}}$. Integrating (\ref{integrand})
over $x$ and $x^{\prime }$ gives a power law dependence of
Im$[\delta f_{\mathrm{ss}}(x_{L},x_{R})]$ on the magnetic field:
\begin{equation}
Im[\delta f_{\mathrm{ss}}(x_{L},x_{R})]\propto
(|H_{z}(y_{1})|^{-3/2}-|H_{z}(y_{2})|^{-3/2})\,,  \label{correct2}
\end{equation}%
where $y_{1}$ and $y_{2}$ are $y$-coordinates of the junction edges.

In contrast to a perpendicular Zeeman field, in a parallel field
$\pm 1$ triplets exponentially decay near boundaries, as can be seen
from Eqs. (\ref{Usadel},\ref{Mz}). So they cannot provide a
long-range link between superconducting terminals.

In conclusion, an analogue to the ISHE exists in DC Josephson SNS
junctions. Unlike the normal ISHE, the supercurrent through the SNS
contact can be induced by a static Zeeman interaction by magnetic or
exchange fields oriented normal to the 2DEG and varying in the
direction transverse to the electric current. A destructive
depairing effect of the strong Zeeman field is diminished by Rashba
SOI leading to a power-low dependence on this field. We show that a
supercurrent through the junction can also be induced by a uniform
parallel Zeeman field, corroborating thus the numerical analysis of
Ref. \cite{Reynoso}. On the other hand, the depairing effect of such
a field was found to be strong (exponential). In both cases an
appearance of the supercurrent can be explained in terms of the
Meissner effect produced by an effective vector potential, which is
a combined effect of the Zeeman field and Rashba spin-orbit
interaction.

We considered the diffusive transport regime which is relevant in
low mobility metals and (magnetic) semiconductors. Furthermore, the
diffusive regime, allows an elucidation of the main physics and
parameters governing this phenomena. We expect a strong Josephson
ISHE in ballistic junctions containing a metallic normal layer with
a strong Rashba interaction, for example in Bi films on some
substrates \cite{Ast}. Ballistic quantum wells of narrow gap
semiconductor are also expected to exhibit an increased Josephson
ISHE.

A.G.M. gratefully acknowledges hospitality of NTNU.

\end{document}